\documentclass[fleqn,usenatbib]{mnras}

\usepackage{newtxtext,newtxmath}

\usepackage[T1]{fontenc}
\usepackage{ae,aecompl}
\usepackage{natbib}

\usepackage{graphicx}	
\usepackage{amsmath}	
\usepackage{amssymb}	
\usepackage{deluxetable}	
\usepackage{adjustbox}
\usepackage{hyperref}
\usepackage{array}

\newcolumntype{L}[1]{>{\raggedright\let\newline\\\arraybackslash\hspace{0pt}}m{#1}}
\newcolumntype{C}[1]{>{\centering\let\newline\\\arraybackslash\hspace{0pt}}m{#1}}
\newcolumntype{R}[1]{>{\raggedleft\let\newline\\\arraybackslash\hspace{0pt}}m{#1}}

\newcommand{\teff}{\ensuremath{T_{\rm eff}}}
\newcommand{\logg}{\ensuremath{\log{g}}}

\newcommand{\feh}{\ensuremath{\rm [Fe/H]}}
\newcommand{\ld}{\ensuremath{\theta_{\rm LD}}}
\newcommand{\ud}{\ensuremath{\theta_{\rm UD}}}
\newcommand{\mm}{\ensuremath{\mu {\rm m}}}
\newcommand{\fbol}{\ensuremath{F_{\rm bol}}}
\newcommand{\ergscmsq}{\ensuremath{\rm erg\,s^{-1}\,cm^{-2}}}

\newcommand{\p}{PIONIER}
\newcommand{\vis}{visibility}
\newcommand{\viss}{visibilities}

\newcommand{\Md}{M-dwarf}
\newcommand{\Mds}{M-dwarfs}
\newcommand{\nir}{near-infrared}

\newcommand{\rsun}{\ensuremath{R_{\sun}}}
\newcommand{\msun}{\ensuremath{M_{\sun}}}
\newcommand{\lsun}{\ensuremath{L_{\sun}}}

\newcommand{\teffsun}{\ensuremath{T_{\mathrm{eff},\sun}}}

\newcommand{\rstar}{\ensuremath{R_\star}}
\newcommand{\mstar}{\ensuremath{M_\star}}
\newcommand{\lstar}{\ensuremath{L_\star}}

\hypersetup{draft}

\title[M-dwarf diameters]{A discontinuity in the \teff-radius relation of M-dwarfs}

\author[Rabus et al.]{%
Markus Rabus,$^{1,2,3,4}$\thanks{mrabus@astro.puc.cl}
R\'egis Lachaume,$^{1,2}$ 
Andr\'es Jord\'an,$^{1,5}$
Rafael Brahm,$^{1,5}$ 
\newauthor
Tabetha Boyajian,$^{6}$
Kaspar von Braun,$^{7}$ 
N\'estor Espinoza,$^{2}$ 
\newauthor
Jean-Philippe Berger,$^{8}$ 
Jean-Baptiste Le Bouquin,$^{9}$
and Olivier Absil,$^{10}$
\\
$^{1}$Instituto de Astronom\'\i{}a, Facultad de F\'\i{}sica, Pontificia Universidad Cat\'olica de Chile, casilla 306, Santiago 22, Chile\\
$^{2}$Max-Planck-Institut f\"ur Astronomie, K\"onigstuhl 17, D-69117 Heidelberg, Germany\\
$^{3}$Las Cumbres Observatory Global Telescope, 6740 Cortona Dr., Suite 102, Goleta, CA 93111, USA\\
$^{4}$Department of Physics, University of California, Santa Barbara, CA 93106-9530, USA\\
$^{5}$Millennium Institute for Astrophysics, Chile\\
$^{6}$Department of Physics and Astronomy, Louisiana State University, 202 Nicholsom Hall, Baton Rouge, LA 70803, USA\\
$^{7}$Lowell Observatory, 1400 W. Mars Hill Road, Flagstaff, AZ 86001, USA\\
$^{8}$Institut de Plan\'etologie et d'Astrophysique de Grenoble, Observatoire de Grenoble, France\\
$^{9}$Department of Astronomy, University of Michigan, 1085 S. University, Ann Arbor, MI 48109, USA\\
$^{10}$STAR Institute, Universit\'e de Li\`ege, 19c All\'ee du Six Ao\^ut, B-4000 Li\`ege, Belgium\\
}

\date{Accepted XXX. Received YYY; in original form ZZZ}

\pubyear{2018}

\begin{document}
\label{firstpage}
\pagerange{\pageref{firstpage}--\pageref{lastpage}}
\maketitle

\begin{abstract}
We report on 13 new high-precision measurements of stellar diameters for low-mass 
dwarfs obtained by means of \nir\, long-baseline interferometry with PIONIER at the 
Very Large Telescope Interferometer. Together with accurate parallaxes from Gaia 
DR2, these measurements provide precise estimates for their linear radii, effective 
temperatures, masses, and luminosities. This allows us to refine the effective 
temperature scale, in particular towards the coolest \Mds. We measure for late-type stars with enhanced metallicity slightly inflated radii, whereas for stars with decreased metallicity we measure smaller radii. We further show that Gaia 
DR2 effective temperatures for \Mds{} are underestimated by $\sim$ 8.2\% and give an  empirical $M_{G}$-\teff\ relation which is better suited for M-dwarfs with \teff\ between 2600 and 4000 K.
Most importantly, we are able to 
observationally identify a discontinuity in the \teff-radius plane, which is likely due to the transition from partially convective \Mds\, to the 
fully convective regime. We found this transition to happen between 3200\,K and 3340\,K, or equivalently for stars with masses $\approx 0.23\,\msun$. We find 
that in this transition region the stellar radii are in the  range from 0.18 to 
0.42\,\rsun{} for similar stellar effective temperatures. 
\end{abstract}




\section{Introduction} \label{sec:intro}

Low-mass dwarfs are the most numerous stars in the Universe, and understanding them is thus clearly an important endeavour. Beyond their own interest, investigations by \citet{bonfils:2013:mdwarfsample}, \citet{dressing:2013:occurencerate} and \citet{kopparapu:2013:occurencerate} have shown that \Mds\, may be the most abundant planet hosts in the Milky Way as well. The estimation of parameters and properties of an exoplanet are intimately 
connected to the stellar host, e.g.\ the stellar mass determines the measured 
semi-amplitude for radial velocity observations and hence influences the mass estimate of 
the planet. In the case of transiting extrasolar planets (TEPs), their physical radii can be measured from the transit shape if the radii of the stellar hosts are known. In addition, the stellar radius and effective temperature are linked to the planet's surface temperature and the location of the habitable zone. All of these examples illustrate
how important stellar astrophysical properties are for the characterization of exoplanets in general. M-dwarfs are attractive targets to search for transiting exoplanets not only due to their numbers, but also due to the fact that for a given planetary size the transit depth is deeper around low-mass stars due to their smaller sizes. Also, the habitable zone around these stars is closer, resulting in shorter periods that make detection easier. Indeed, one of the main drivers for the upcoming TESS mission \citep{TESS} is to detect transiting exoplanets around low-mass stars.

A fundamental stellar property is the radius, and for low-mass stars its estimation has been done mostly through stellar models. Fortunately, considerable improvements in interferometric observation techniques allow us now to obtain stellar parameters such as the stellar radius directly. However, these measurements become more difficult as we go to cooler dwarf stars due to their inherently lower luminosity and smaller radii. Measured angular diameters of \Mds\ are generally close to the current baseline limit of available interferometers. Up to now, extensive interferometric observations on \Md\ stars have been done mainly from the Northern Hemisphere with the CHARA array  \citep{ berger:2006, braun:2011:gj581, boyajian:2012:mdwarfs, braun:2014:diameters} and a few with the VLT-Interferometer (VLTI) from the South \citep{ segransan:2003, demory:2009}. These interferometric direct measurements showed a discrepancy with the parameters measured indirectly \citep{boyajian:2012}. The work of \citet{boyajian:2012} found in particular large disagreements for low-mass stars, where the radii measured by interferometers were more than 10\% larger than the ones based on models from \citet{chabrier:1997:stellarstructure}. Likewise, \citet{Kesseli:2018} found that this inflation of the M-dwarf radii extends down to the fully convective regime with a discrepancy of 13\% -- 18\%.

This discrepancy affects in turn other stellar parameters like surface temperature (\teff), gravities (\logg), masses, luminosities, and eventually also possible planetary parameters. Therefore, it is important to observe and re-evaluate the properties of more \Md\, stars with interferometric observations, particularly towards the later spectral types which have not been extensively studied at all.

Theoretical stellar evolution  models for low-mass stars predict a transition into the fully convective regime to occur somewhere between 0.2\,\msun{} \citep{dorman1989} and 0.35\,\msun{} \citep{chabrier:1997:stellarstructure}, depending on the underlying stellar model. For partially convective stars, the stellar structure is Sun-like, having a radiative zone and a convective envelope. The only previous observational indications for this transition in late-type stars are based on observations of magnetic fields and measurements of stellar rotational periods. \citet{browning:2008} showed that stars whose convective region extends to the core have strong large-scale magnetic fields and, in fact, we have observational evidence that the fraction of \Mds{} with strong magnetic fields on a large scale is higher for mid- to late-type \Mds{} than for early type ones \citep{donati:2008}. On the other hand, \citet{wright:2016} showed that rotation-dependent dynamos are very similar in both partially and fully convective stars. \citet{irwin:rotation} and \citet{newton:rotation} measured rotational spin-velocities of M-stars. The authors found two divergent populations of faster and slower rotators in the fully convective mass regime, which makes rotation measurements difficult to use in the determination of whether a late-type star is fully convective. Moreover, the rotation of fully convective stars depends on both age and mass. All former indications of fully convective stars have been done indirectly and are not unambiguous.

In this work we present directly measured physical parameters for a sample of 13 low-mass stars using observations with the VLT-Interferometer (VLTI). These observations are used to probe the transition between the partially and fully convective regimes and to identify the dependence of the stellar radii on other stellar properties. The paper is structured as follows. In \S\ref{sec:obs} we lay out the observational details. In \S\ref{sec:phys_params} we detail how we estimated the stellar physical parameters. Finally, we discuss the implication of the measured stellar parameters on stellar evolution and structure models in \S\ref{sec:discussion} and we conclude in \S\ref{sec:conclusion}.

\section{Observations and Data Reduction} 
\label{sec:obs}

\subsection{\p{} observations} 
\label{subsec:pionierobs}

Our target sample is compiled from a list of \Mds{} within $\sim 15$\,pc (so the stars are resolved within the given VLTI baseline) and with H-band magnitudes $<7$ (so that fringes will be easily visible and we can obtain a good signal-to-noise ratio). 

In order to measure the angular diameter of our sample stars, we used the VLTI/\p{} interferometer \citep{pionier}. \p{} is an integrated optics four-beam combiner operating at the \nir{} wavelength range. We used the auxiliary telescopes (ATs) in a A1-G1-K0-J3 quadruplet configuration. This configuration gave us the longest VLTI baseline available (from 57 meters between the stations K0 and J3, up to 140 metres between A1 and J3) and we used the Earth's rotation to further fill the $(u, v)$ plane.

We observed our sample with a three-channel spectral dispersion (SMALL mode), whenever possible. In cases where this was not possible, due to low coherence time on a given night or the relative faintness of the target, we observed without spectral dispersion (FREE mode). Similarly, the number of scan steps were adjusted according to the objects' brightness and atmospheric conditions. As our sample stars were not too bright we were able to use the fast Fowler readout mode for all of our observations.

Our observing strategy was to bracket each science frame (SCI) with a calibrator star (CAL), observed with the same setup as the science object. The calibrators are chosen to be mostly point-like nearly unresolved stars \citep{vanbelle:calibrators}, so the uncertainties in their diameter will not influence our targets, but we also included calibrators with known diameter for verification proposes. We also made sure that the visibility precision of our calibrators was below 1\%. In order to search for suitable calibrators, we used the ASPRO2-tool and SearchCal\footnote{\url{http://www.jmmc.fr/aspro\_page.htm}}. For each science target we repeated around 11 times a CAL-SCI-CAL block, and in each block we used different calibrator stars. The same target was also observed on different nights. This strategy helped us to beat down the systematic noise from the instrument and atmosphere. We reduced our observed raw fringes to calibrated visibilities and closure phases with a modified version of the \p{} data reduction software \citep[\emph{pndrs}, described in][]{Lachaume:2019}. 

\subsection{Calibrated Visibilities and Angular diameters} \label{subsec:calvis}

Our modified data reduction with \emph{pndrs} is fully described in detail in a publication by \citet[]{Lachaume:2019}, where we also show a rigorous analysis of the interferometric measurement errors. Here we will give only a brief summary of the data reduction process and we refer interested readers to \citet[]{Lachaume:2019}, for more details on the data analysis. In the first step we calibrate the detector frames. This was done by dark correcting the detector data and from the kappa-matrix we calibrated the transmission of the respective baseline. Finally, we used frames illuminated by an internal light source to calibrate the wavelength. Basically, these calibrated frames will allow us to obtain the raw visibilities, which are in turn the product of true \viss{} and the system transfer function. The system transfer function characterizes the response of the interferometer as a function of spatial frequency and in order to get the true \viss{} it needs to be estimated by using calibrator stars. Assuming that all our calibrator stars have well known true \viss, i.e. an unresolved calibrator has a known \vis{} of unity and a resolved star has a known diameter, either measured or from spectral typing. By further assuming a smooth transfer function, in theory this would allow us to calibrate our raw \viss. Nevertheless, uncertainties in the assumed calibrators' diameters can impact all observations in a sequence due to systematic errors in the transfer function estimate \citep[][and references therein]{Lachaume:2019}. Further errors can be introduced through systematic uncertainties in the absolute wavelength calibration \citep{Gallenne:wavelength_cal:2018} and by several other random effects which will affect the different spectral channels in a similar or imbalanced manner, like e.g. atmospheric jitter or flux variations between the arms of the interferometer. In order to account for the correlation effects in our observations we apply a bootstrap method as described in \citet{Lachaume:2019,bootstrap}. 

\begin{figure}
\includegraphics[scale=0.27]{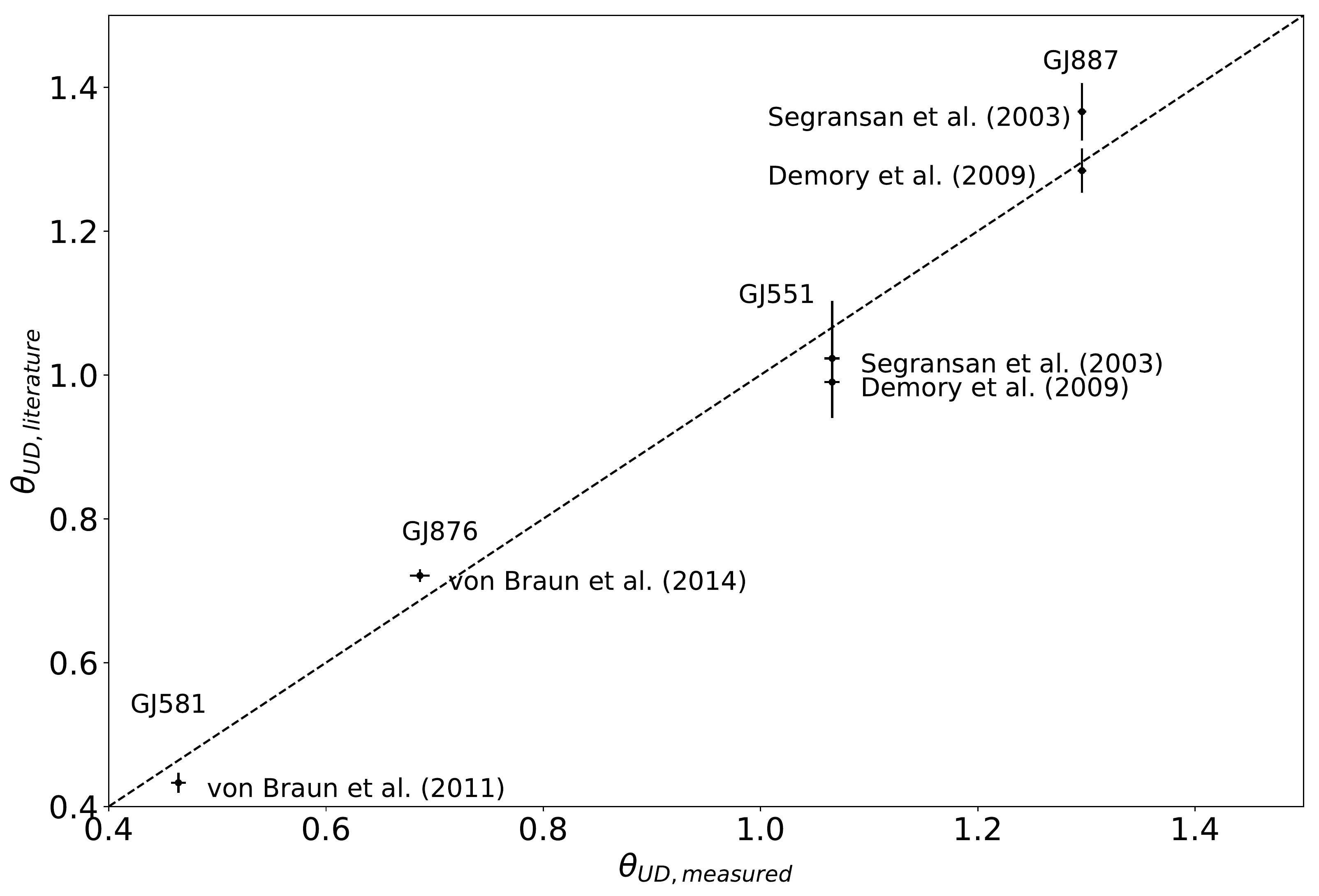}
\caption{Our angular diameters \ud\, compared to literature values. We find good agreement between our measurements and the literature.
\label{fig:compareUD}}
\end{figure}

Generally, in a bootstrap one resamples several times new data sets from the empirical data itself by replacing parts of the original data. For each candidate, we started by picking randomly interferograms out of the parent population of $\sim 10^2$ interferograms. These interferograms are reduced and averaged to a single data set, which corresponds to the raw visibility. As mentioned before, uncertainties in the calibrators' diameter can cause correlated errors. Therefore, we choose arbitrarily a calibrator with a diameter, drawn randomly from a Gaussian distribution centered on the catalogued diameter and with a width corresponding to the error bars. We used 6 to 18 data sets and calibrators to replace the original data and to calculate the system transfer function and calibrated visibilities. We repeat this procedure to obtain 5,000 bootstrap realizations. These calibrated visibilities were fitted with a uniform disk (\ud) model to obtain a distribution of angular diameters for each star observed. In Fig.~\ref{fig:compareUD} we compare some of our measured \ud{} with the ones available in the literature. We find a good agreement between our measurements and the literature values.

\section{Estimating the Physical Parameters from Interferometry} \label{sec:phys_params}

\subsection{Calculation of the stellar radius}

The limb darkened disk \ld{} is usually obtained by fitting directly a limb darkened disk model to the squared \viss, assuming a certain limb darkening law and coefficient. Generally, a linear limb darkening law is assumed and tabulated values are used for the coefficients, see e.g.\ \citet{boyajian:2012}, \citet{braun:2014:diameters}, and \citet{braun:2011:gj581}. We note, that while \ud s are independent of stellar models, photospheric diameters, \ld s depend on stellar models as the limb-darkening coefficient are derived from them. However, the impact on the radius estimate by the limb-darkening in the \nir{} is small (2--4\%) and it is mostly dominated by the angular diameter measurement uncertainties and systematics. 

In order to estimate the \ld, we used the \ud--\ld{} relation from \citet{hanbury_LD}:

\begin{equation}
\ld(\lambda)= \ud\, \sqrt[]{\frac{1-\frac 13\mu(\lambda,\teff,\logg)}{1-\frac7{15}\mu(\lambda,\teff,\logg)}   },
\end{equation}
where \ud{} is the angular diameter we obtained from the calibrated \viss{} and $\mu_\lambda$ is the linear limb darkening coefficient as function of wavelength, \teff{} and \logg. Rather than using tabulated coefficient, we calculated a grid of limb darkening coefficients following \citet{espinoza:2015} corresponding to the atmosphere grid with \teff{} in range 2300--4500\,K, \logg{} in range 4.0--6.0 and a fixed metallicity of 0.0. This allows us to have a conformity with the grid which will be used in Sect. \ref{fbol_estimate}. As filter transmission function of PIONIER, we used a top hat function between 1.5\mm{} and 1.8\mm. 

\subsection{\teff{} estimate}

The measured diameters can be related to the effective temperature by 

\begin{equation}
\teff = \sqrt[4]{\frac{4\fbol}{\sigma\ld}}, 
\end{equation}
where \fbol{} is the bolometric flux (obtained by e.g. fitting the spectral energy distribution with literature photometry to spectral templates), \ld{} is the limb darkened angular diameter and $\sigma$ is the Stefan-Boltzmann constant. 

\subsection{Bolometric flux estimate}
\label{fbol_estimate}

In order to estimate the bolometric flux we started by using the PHOENIX atmosphere models from \citet{husser} to create a grid of synthetic photometric points for filters with available photometric observations of our sample stars. Their models are defined in the wavelength range from 0.05 to 5.5\mm. Our flux model grid runs \teff{} from 2300\,K to 4500\,K, \logg{} between 4.0 and 6.0 dex, and for a fixed metallicity of 0.0 dex. The flux was integrated over the respective band and convolved with the filter profiles from \citet{filtertransmission}. We linearly interpolated this grid of synthetic flux in-between.
The bolometric flux \fbol\, of a given star is then defined as:

\begin{equation}
\fbol = \int_0^{+\infty} F_\mathrm{model}(\lambda,\teff,\logg) \frac{\rstar^2}{d^2} \mathrm{d}\lambda,
\end{equation}
where \rstar{} is the stellar radius and $d$ is the distance. 

\subsection{Multinest fitting for \teff, \rstar, and \lstar}
\label{subsec:multinest}

We first collected observed fluxes for our stars using the VizieR photometric query. To these observed fluxes we fitted the model grid using the \emph{pymultinest} code \citep{pymultinest}. This program is a python code for multimodal nested sampling technique \citep{nestedsampling, multinest}. Our log-likelihood function is

\def\Fobs{\ensuremath{F_{i,\mathrm{obs}}}}
\def\Fmodel{\ensuremath{F_{i,\mathrm{mod}}}}
\def\Nphot{\ensuremath{N_\mathrm{phot}}}

\begin{equation}
\mathcal{\log L}= -\sum_{i=1}^{\Nphot}
	\left[
   		\frac{(\Fobs - \Fmodel)^2}{2\sigma_i^2} 
       - \log{ \frac 1 {\sigma_i\sqrt{2\pi}} }
    \right],
\end{equation}
where $\Fobs$ is the observed flux in a given filter $i$, $\Fmodel$ is the synthetic flux in that filter obtained from the atmosphere models, and $\sigma_i$ is the corresponding measurement error of the observed flux. The sum goes over the $\Nphot$ photometric measurements of a given star.

\begin{figure}
\includegraphics[scale=0.67]{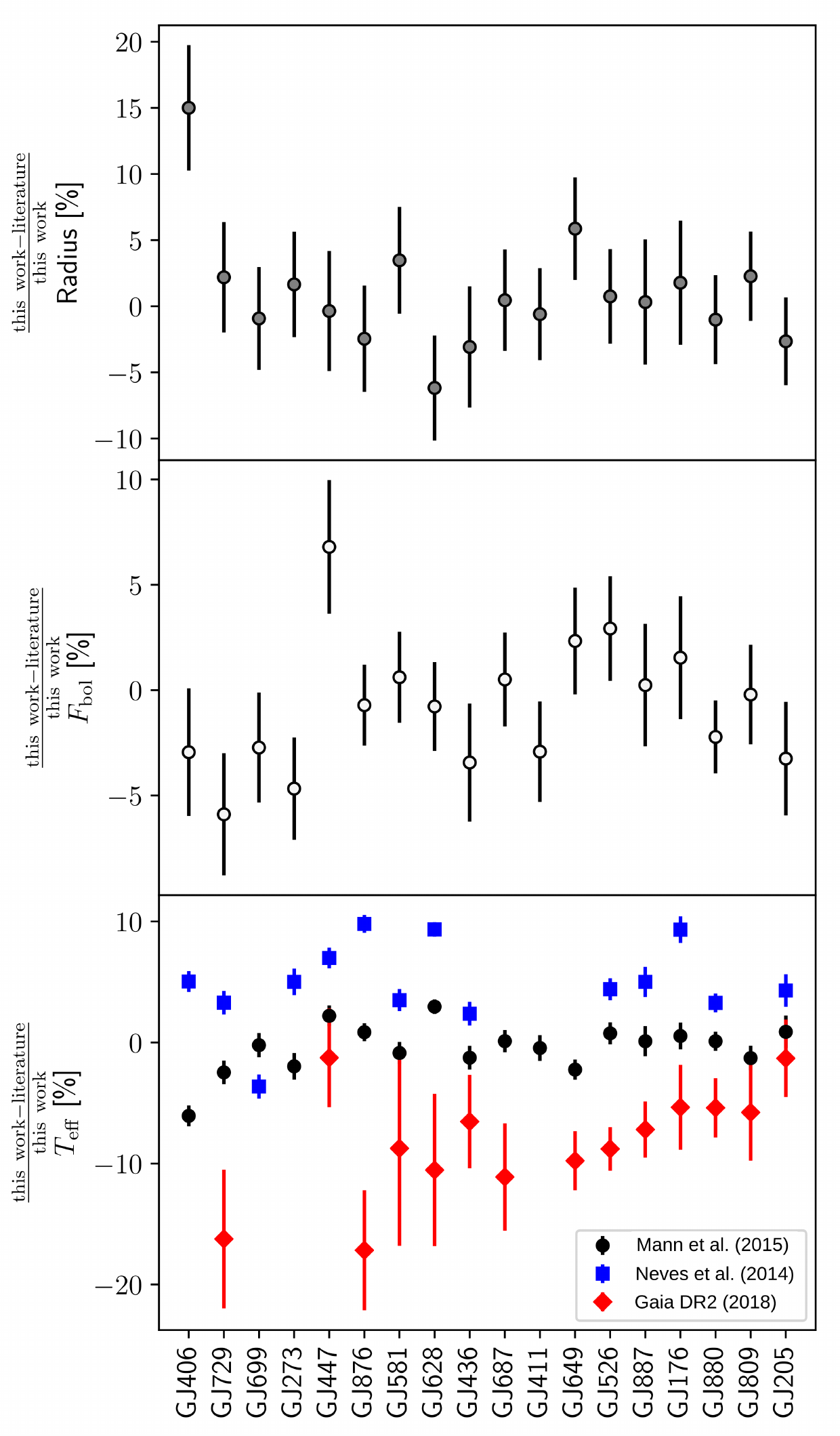}
\caption{We compare our calculated radius, \fbol, and \teff{} with the ones from \citet{mann:2015}. Stars are ordered according to our calculated \teff{} from low (left) to high (right) temperature. The difference between the estimates is small, the mean difference for the radius is 2.9\%, \fbol{} is 2.5\% and \teff{} is 1.4\%. However, by comparing our \teff{} with the ones from optical spectroscopy \citep{neves:2014:feh}, we find an higher mean difference of 5.4\% and -8.2\% for Gaia DR2 \teff. However, single \teff s from Gaia DR2 can have differences of up to $\sim 15 \%$. (See Sect. \ref{subsec:multinest} for details)
\label{fig:comparelit}}
\end{figure}

Our priors are \teff, \logg, distance, and angular diameter \ud. All our priors were drawn from a normal distribution centered at the literature value and with a dispersion corresponding to the respective error. We further repeated this process using \Mds{} with measured diameters from \citet{braun:2012:gj436}, \citet{boyajian:2012:mdwarfs} and \citet{braun:2014:diameters}. Our final parameter estimates are shown in Table \ref{tab:multinest_results}. We compare our values with the ones from \citet{mann:2015} in Figure~\ref{fig:comparelit} and find good agreement with a mean difference of 3\% for all three parameters (from top to bottom: radius, \fbol, \teff). In the same Figure~\ref{fig:comparelit} (bottom plot), we further compare our effective temperatures with the ones obtained by \citet{neves:2014:feh} through spectral type classification using optical spectroscopy and from Gaia DR2 using Apsis \citep{gaiadr2_teff:2018}. In the latter cases the relative difference for \teff{} is generally higher, with a mean difference of 5.4\% and $-8.2\%$ respectively. Therefore, spectral typing of \Mds{} in the optical wavelength range generally overestimates \teff s, whereas Gaia DR2 \teff s are considerably underestimated.

\begin{table*}
\caption{Final parameter estimates obtained through multi-modal nested sampling technique. (See Sect. \ref{subsec:multinest} for details)}
\label{tab:multinest_results}
\begin{tabular}{cccccccc}
\hline \hline
star &  \ud   & \ld  & $\mu_{\lambda}$ & \fbol & \rstar & parallax & calculated \\
name & [mas] & [mas] &   & [$10^{-8}$ \ergscmsq] & [\rsun] & [mas] & \teff [K]\\ 
\hline
GJ~1   & $0.794\pm 0.005$ & $0.812\pm 0.005$ & 0.290 & $3.751\pm 0.072$ & $0.379\pm 0.002$ & $230.133\pm 0.059$ & $3616\pm 14$ \\
GJ~273 & $0.763\pm 0.010$ & $0.783\pm 0.010$ & 0.335 & $2.288\pm 0.118$ & $0.320\pm 0.005$ & $262.961\pm 1.387$ & $3253\pm 39$ \\
GJ~406 & $0.562\pm 0.020$ & $0.582\pm 0.020$ & 0.449 & $0.563\pm 0.044$ & $0.159\pm 0.006$ & $394.867\pm 7.893$ & $2657\pm 20$ \\
GJ~447 & $0.524\pm 0.029$ & $0.540\pm 0.029$ & 0.365 & $1.103\pm 0.091$ & $0.196\pm 0.010$ & $296.309\pm 0.069$ & $3264\pm 24$ \\
GJ~551 & $1.066\pm 0.007$ & $1.103\pm 0.007$ & 0.422 & $2.866\pm 0.210$ & $0.154\pm 0.001$ & $768.500\pm 0.203$ & $2901\pm 68$ \\
GJ~581 & $0.464\pm 0.007$ & $0.476\pm 0.007$ & 0.324 & $0.967\pm 0.039$ & $0.322\pm 0.005$ & $158.747\pm 0.051$ & $3366\pm 28$ \\
GJ~628 & $0.644\pm 0.014$ & $0.661\pm 0.014$ & 0.335 & $1.882\pm 0.068$ & $0.306\pm 0.007$ & $232.209\pm 0.063$ & $3372\pm 12$ \\
GJ~674 & $0.720\pm 0.037$ & $0.737\pm 0.037$ & 0.318 & $2.443\pm 0.232$ & $0.360\pm 0.018$ & $219.800\pm 0.047$ & $3409\pm 25$ \\
GJ~729 & $0.625\pm 0.020$ & $0.642\pm 0.020$ & 0.345 & $1.370\pm 0.096$ & $0.205\pm 0.006$ & $336.121\pm 0.064$ & $3162\pm 30$ \\
GJ~832 & $0.794\pm 0.010$ & $0.814\pm 0.010$ & 0.325 & $3.359\pm 0.113$ & $0.435\pm 0.005$ & $201.407\pm 0.043$ & $3512\pm 23$ \\
GJ~876 & $0.686\pm 0.009$ & $0.705\pm 0.009$ & 0.342 & $1.902\pm 0.058$ & $0.354\pm 0.005$ & $213.866\pm 0.078$ & $3275\pm 18$ \\
GJ~887 & $1.297\pm 0.005$ & $1.328\pm 0.004$ & 0.323 & $10.916\pm 0.657$ & $0.470\pm 0.001$ & $304.219\pm 0.044$ & $3692\pm 57$ \\
\hline
\multicolumn{2}{l}{Literature stars}\\
\hline
GJ~176 & $0.442\pm 0.020$ & $0.452\pm 0.020$ & 0.306 & $1.274\pm 0.099$ & $0.460\pm 0.020$ & $105.565\pm 0.069$ & $3700\pm 45$ \\
GJ~205 & $0.904\pm 0.003$ & $0.924\pm 0.003$ & 0.283 & $6.140\pm 0.400$ & $0.566\pm 0.002$ & $175.430\pm 0.069$ & $3835\pm 69$ \\
GJ~411 & $1.380\pm 0.013$ & $1.412\pm 0.013$ & 0.301 & $10.514\pm 0.515$ & $0.387\pm 0.004$ & $392.630\pm 0.675$ & $3547\pm 40$ \\
GJ~436 & $0.405\pm 0.013$ & $0.415\pm 0.013$ & 0.321 & $0.800\pm 0.053$ & $0.436\pm 0.013$ & $102.497\pm 0.093$ & $3436\pm 33$ \\
GJ~526 & $0.807\pm 0.013$ & $0.824\pm 0.013$ & 0.287 & $4.134\pm 0.183$ & $0.482\pm 0.008$ & $183.983\pm 0.051$ & $3677\pm 30$ \\
GJ~649 & $0.472\pm 0.012$ & $0.483\pm 0.012$ & 0.294 & $1.329\pm 0.072$ & $0.539\pm 0.013$ & $96.314\pm 0.031$ & $3619\pm 25$ \\
GJ~687 & $0.830\pm 0.013$ & $0.850\pm 0.013$ & 0.317 & $3.380\pm 0.145$ & $0.416\pm 0.007$ & $219.781\pm 0.033$ & $3443\pm 29$ \\
GJ~699 & $0.917\pm 0.005$ & $0.941\pm 0.005$ & 0.342 & $3.176\pm 0.120$ & $0.185\pm 0.001$ & $548.358\pm 1.513$ & $3221\pm 32$ \\
GJ~809 & $0.698\pm 0.008$ & $0.715\pm 0.008$ & 0.314 & $3.341\pm 0.148$ & $0.541\pm 0.006$ & $142.033\pm 0.030$ & $3743\pm 39$ \\
GJ~880 & $0.716\pm 0.004$ & $0.736\pm 0.004$ & 0.357 & $3.468\pm 0.084$ & $0.544\pm 0.003$ & $145.610\pm 0.038$ & $3724\pm 23$ \\
\hline
\end{tabular}
\end{table*}

\subsection{Mass estimates}
\label{subsec:mass_estimate}
The mass cannot be measured directly from interferometry. Therefore, we make use of a fully empirical model-independent mass-luminosity relation (MLR) from \citet{MLrelation:benedict} and \citet{Mann:MLR:2018}. In both cases we use their calibration relations in K-band, therefore
for all our stars we collected SAAO K-band magnitudes from \citet{koen:2010} and Ks-band magnitudes from \citet{mann:2015} and \citet{2003tmc..book.....C}. The corresponding magnitudes are given in Table \ref{tab:mass}. The SAAO K-band magnitudes were transformed to 2MASS Ks using the transformation\footnote{\url{http://http://www.astro.caltech.edu/~jmc/2mass/v3/transformations/}} $ {\rm Ks_{2MASS}}	=	{\rm K_{SAAO}} - (0.024 \pm 0.003)	+	(0.017 \pm 0.006) {\rm  (J-K)}_{SAAO}$.

We converted the Ks-band magnitudes to absolute magnitudes using the respective parallax given in Table \ref{tab:multinest_results} and estimated the mass for a given star. 
From the mass and radius, we were also able to calculate the surface gravity (\logg): 
\begin{equation}
g_\star=\frac{G\mstar}{\rstar^2}, 
\end{equation}
where $G$ is the gravitational constant. Table \ref{tab:mass} shows a summary of the calculated mass, luminosity and \logg\, for our sample.

\begingroup
\renewcommand*{\thefootnote}{\alph{footnote}}
\begin{table*}
\caption{Calculated other distance-dependent stellar parameters. (See Sect. \ref{subsec:mass_estimate})}
\label{tab:mass}
\begin{tabular}{rccccc}
\hline \hline
\multicolumn{1}{c}{star} name & Ks [mag.] & M$_{\rm Ks}$ [mag.] & \mstar [\msun]\footnotemark[1] & \lstar [\lsun] & \logg [dex] \\
\hline
GJ1   & 4.53$\pm $0.01\footnotemark[2] & 6.33$\pm $0.01 & 0.390$\pm $0.010 & 0.0220$\pm $0.0004 & 4.87 \\
GJ176 & 5.63$\pm $0.01\footnotemark[2] & 5.74$\pm $0.01 & 0.486$\pm $0.011 & 0.0356$\pm $0.0028 & 4.80 \\
GJ205 & 3.86$\pm $0.02\footnotemark[3] & 5.08$\pm $0.02 & 0.590$\pm $0.015 & 0.0621$\pm $0.0041 & 4.70 \\
GJ273 & 4.87$\pm $0.01\footnotemark[2] & 6.97$\pm $0.02 & 0.293$\pm $0.007 & 0.0103$\pm $0.0005 & 4.89 \\
GJ406 & 6.15$\pm $0.02\footnotemark[3] & 9.13$\pm $0.05 & 0.110$\pm $0.003 & 0.0011$\pm $0.0001 & 5.08 \\
GJ411 & 3.36$\pm $0.02\footnotemark[3] & 6.33$\pm $0.02 & 0.390$\pm $0.010 & 0.0212$\pm $0.0010 & 4.85 \\
GJ436 & 6.04$\pm $0.02\footnotemark[3] & 6.09$\pm $0.02 & 0.429$\pm $0.010 & 0.0237$\pm $0.0016 & 4.79 \\
GJ447 & 5.68$\pm $0.02\footnotemark[3] & 8.04$\pm $0.02 & 0.174$\pm $0.004 & 0.0039$\pm $0.0003 & 5.09 \\
GJ526 & 4.56$\pm $0.02\footnotemark[3] & 5.89$\pm $0.02 & 0.463$\pm $0.011 & 0.0380$\pm $0.0017 & 4.74 \\
GJ551 & 4.38$\pm $0.03\footnotemark[4] & 8.81$\pm $0.03 & 0.124$\pm $0.003 & 0.0015$\pm $0.0001 & 5.16 \\
GJ581 & 5.85$\pm $0.01\footnotemark[2] & 6.85$\pm $0.01 & 0.310$\pm $0.007 & 0.0119$\pm $0.0005 & 4.91 \\
GJ628 & 5.09$\pm $0.01\footnotemark[2] & 6.92$\pm $0.01 & 0.300$\pm $0.007 & 0.0109$\pm $0.0004 & 4.94 \\
GJ649 & 5.63$\pm $0.02\footnotemark[3] & 5.55$\pm $0.02 & 0.517$\pm $0.013 & 0.0446$\pm $0.0024 & 4.69 \\
GJ674 & 4.86$\pm $0.01\footnotemark[2] & 6.57$\pm $0.01 & 0.352$\pm $0.008 & 0.0157$\pm $0.0015 & 4.87 \\
GJ687 & 4.50$\pm $0.02\footnotemark[3] & 6.21$\pm $0.02 & 0.409$\pm $0.010 & 0.0218$\pm $0.0009 & 4.81 \\
GJ699 & 4.53$\pm $0.02\footnotemark[3] & 8.23$\pm $0.02 & 0.160$\pm $0.004 & 0.0033$\pm $0.0001 & 5.11 \\
GJ729 & 5.39$\pm $0.02\footnotemark[3] & 8.03$\pm $0.02 & 0.175$\pm $0.004 & 0.0038$\pm $0.0003 & 5.06 \\
GJ809 & 4.58$\pm $0.02\footnotemark[3] & 5.34$\pm $0.02 & 0.551$\pm $0.014 & 0.0515$\pm $0.0023 & 4.71 \\
GJ832 & 4.46$\pm $0.01\footnotemark[2] & 5.98$\pm $0.01 & 0.447$\pm $0.011 & 0.0258$\pm $0.0009 & 4.81 \\
GJ876 & 5.04$\pm $0.01\footnotemark[2] & 6.69$\pm $0.01 & 0.334$\pm $0.008 & 0.0129$\pm $0.0004 & 4.86 \\
GJ880 & 4.54$\pm $0.02\footnotemark[3] & 5.36$\pm $0.02 & 0.547$\pm $0.014 & 0.0509$\pm $0.0012 & 4.71 \\
GJ887 & 3.33$\pm $0.02\footnotemark[3] & 5.74$\pm $0.02 & 0.486$\pm $0.012 & 0.0367$\pm $0.0022 & 4.78 \\ 
\hline
\end{tabular}
\tablerefs{
\footnotemark[1] Estimated using MLR from \citet{Mann:MLR:2018}
\footnotemark[2] \citet{koen:2010};
\footnotemark[3] \citet{mann:2015};
\footnotemark[4] \citet{cutri:2003:2mass}}
\end{table*}
\endgroup


\section{Discussion} \label{sec:discussion}

\begin{figure}
\includegraphics[scale=0.23]{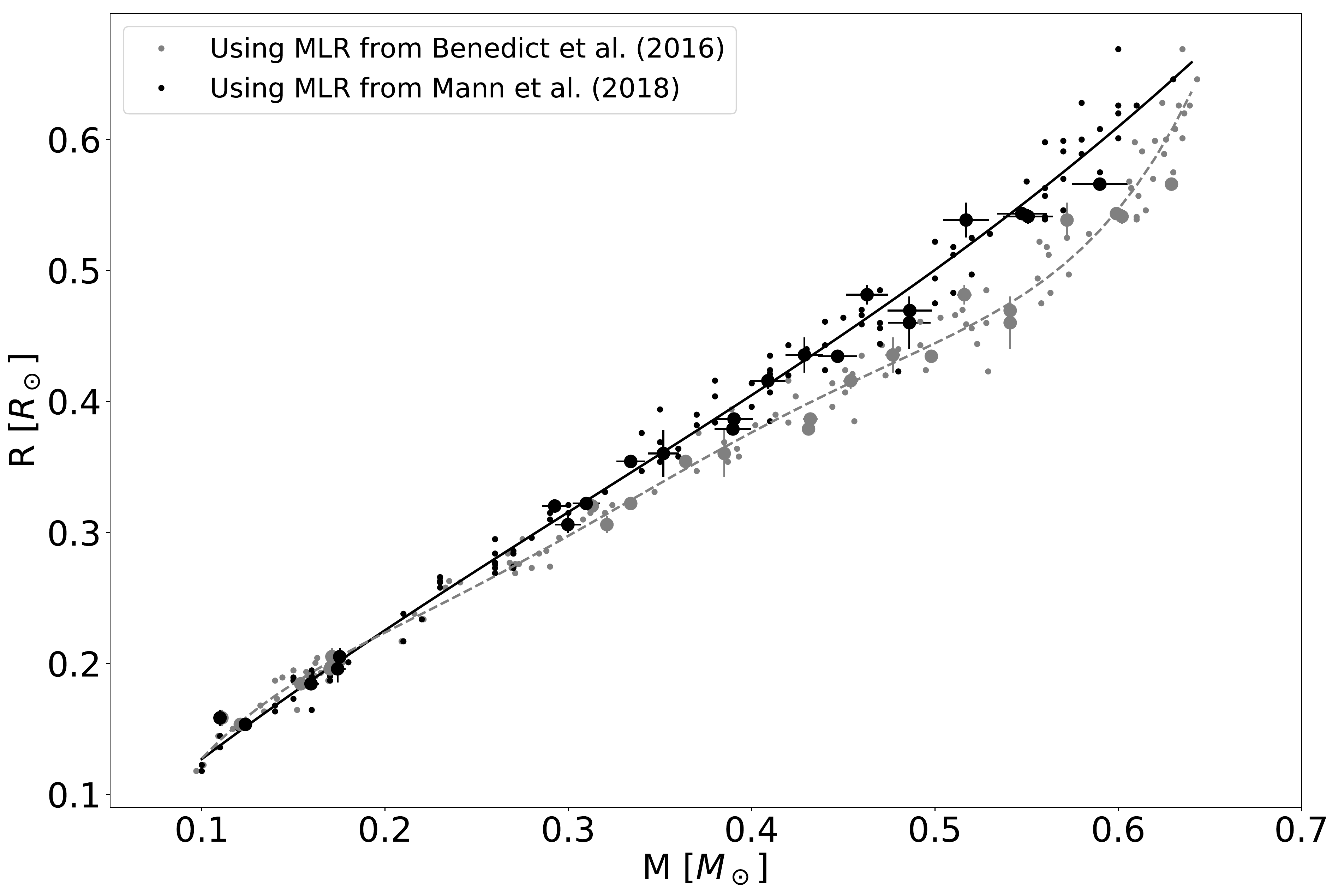}
\caption{Radius-mass relation for our sample (large dots) and a subset of data from \citet[][small dots]{mann:2015}, selected as described in the text. Black dots are based on mass estimates using the MLR from \citet{Mann:MLR:2018}, whereas grey dots are based on the one from \citet{MLrelation:benedict}. The lines show best-fit polynomial, as resulted from the respective MLR.  (Details are discussed in Sect. \ref{sec:discussion})
\label{fig:MR-relation}}
\end{figure}

In order to discuss the behaviour of our sample stars we investigate some relations between the available parameters. In the following analysis we also added stars from \citet{mann:2015} which have measured Gaia DR2 parallaxes \citep{gaiaDR2}. In order to avoid contamination, the population from \citet{mann:2015} has further been cleaned by removing double stars and variable stars (as, e.g., BY-Dra type). We start by constructing a relation between the stellar radius and stellar mass (MR-relation), shown in Figure~\ref{fig:MR-relation}. As pointed out by \citet{Mann:MLR:2018}, comparing their MLR to the one from \citet{MLrelation:benedict} resulted in a discrepancy of more than 10\% for stars with masses $>$0.3\msun. This discrepancy is also visible in Figure~\ref{fig:MR-relation}, where the black dots represents the masses calculated using the MLR relation from \citet{Mann:MLR:2018} and the grey dots using \citet{MLrelation:benedict}. Above 0.3\msun\ we get higher masses for the same star using \citet{MLrelation:benedict}, compared to \citet{Mann:MLR:2018}.
We also fitted polynomials of different degrees to each relation using the Levenberg-Marquardt algorithm. For each polynomial, we calculated the Akaike information criterion \citep[AIC]{akaike:1974} and the Bayesian Information criterion \citep[BIC]{schwarz197801}. We found, that by using the MLR relation from \citet{Mann:MLR:2018}, the best-fit polynomial for the MR relation is of 3rd order, whereas by using \citet{MLrelation:benedict}, it is a 5th order polynomial. The high order structure caused by the MLR from \citet{MLrelation:benedict} is also visible in Figure~\ref{fig:MR-relation}. Since \citet{Mann:MLR:2018} has been calibrated using accurate Gaia DR2 parallaxes, we continue to use their relation. We find that in this case the mass-radius relation is best characterized by a cubic order polynomial of the form:

\begin{equation}
\begin{split}
\frac\rstar\rsun = & 0.013(\pm 0.010) 
           +1.238(\pm 0.117)\,\frac\mstar\msun \\
         &-1.13(\pm 0.40)\,\left(\frac\mstar\msun\right)^2 
          +1.21(\pm 0.42)\,\left(\frac\mstar\msun\right)^3 \\
\end{split}
\end{equation}

The standard deviation of the residuals is $0.016\,\rsun$ and the median absolute deviation (MAD) $0.008\,\rsun$. The errors of the polynomial coefficients (closed brackets) are estimated from the covariance matrix. 

\begin{figure*}
\includegraphics[scale=0.47]{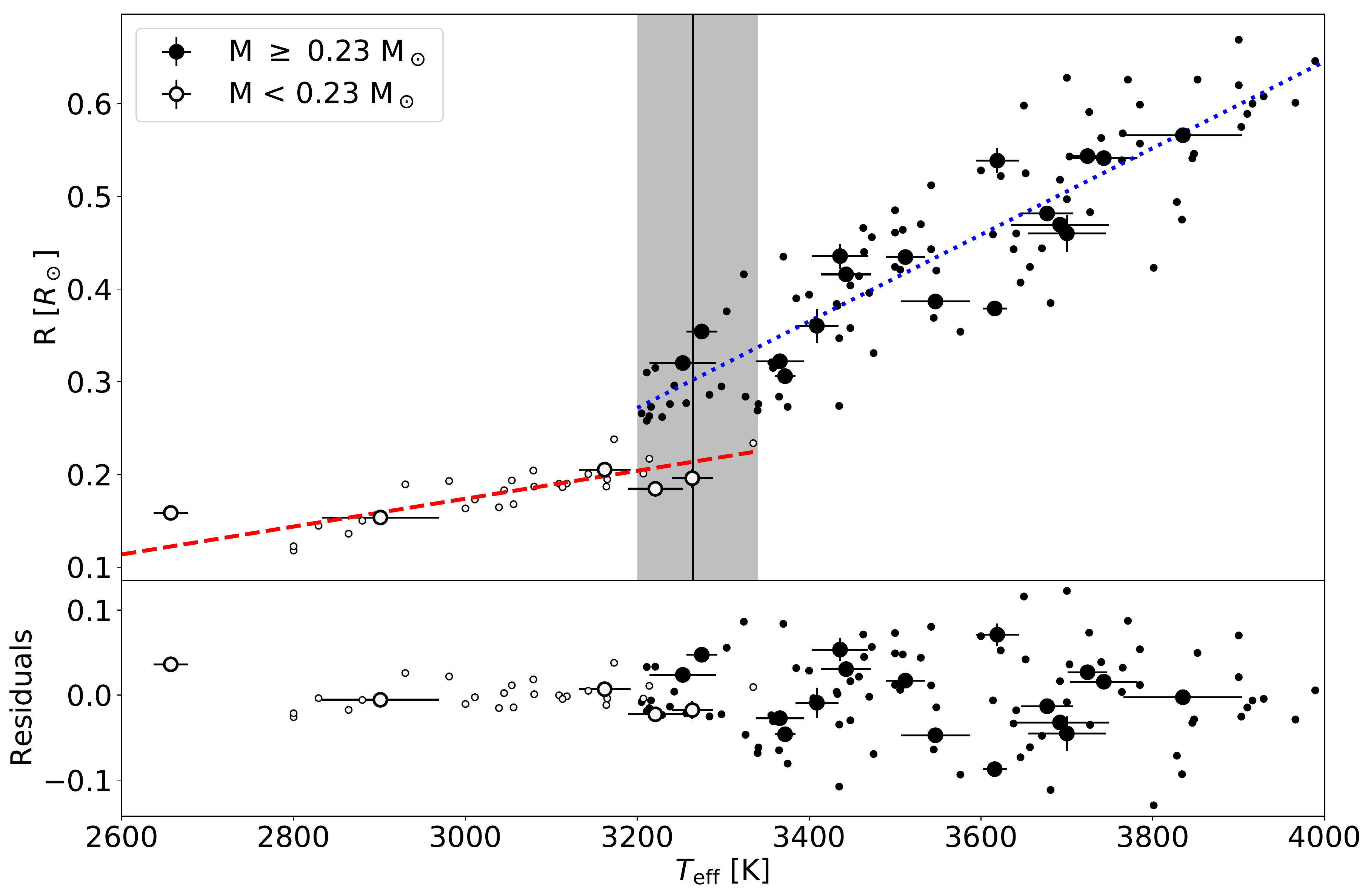}
\caption{ \teff{} versus radius plot with the stellar mass coded as filled and empty circles, respectively. The lines show fitted polynomials for the two different mass populations (red dashed line for stars with masses $< 0.23\,\msun$ and blue dotted line for stars with masses $\geq 0.23\,\msun$). Grey shaded area shows the region where we find a possible discontinuity and which we attribute to the transition between partially and fully convective stars. We also added \Mds{} from \citet{mann:2015} (small dots), see text for details. The lower plot shows the residual after subtracting the polynomials in Eq. \ref{eq:RTeff} from the radius measurements. (Details are discussed in Sect. \ref{sec:discussion})
\label{fig:teffR}}
\end{figure*}

We also establish a relation between the stellar radius and its effective temperature, see Figure~\ref{fig:teffR}. Interestingly, in Figure~\ref{fig:teffR} we identified a discontinuous behaviour between 3200 and 3340 K (gray shaded area), where the radius spans a range from 0.18 to 0.42 \rsun\, for similar effective temperatures. Considering that our mean measurement error for the radius is $\sim 0.006\,\rsun$, this corresponds to a 40-$\sigma$ difference. We also note that we have done a detailed error analysis of our diameter measurements in \citet{Lachaume:2019}. We further find that this discontinuity corresponds to a mass of 0.23\,\msun, see filled and empty dots in Figure~\ref{fig:teffR}. 

\begin{figure*}
\includegraphics[scale=0.47]{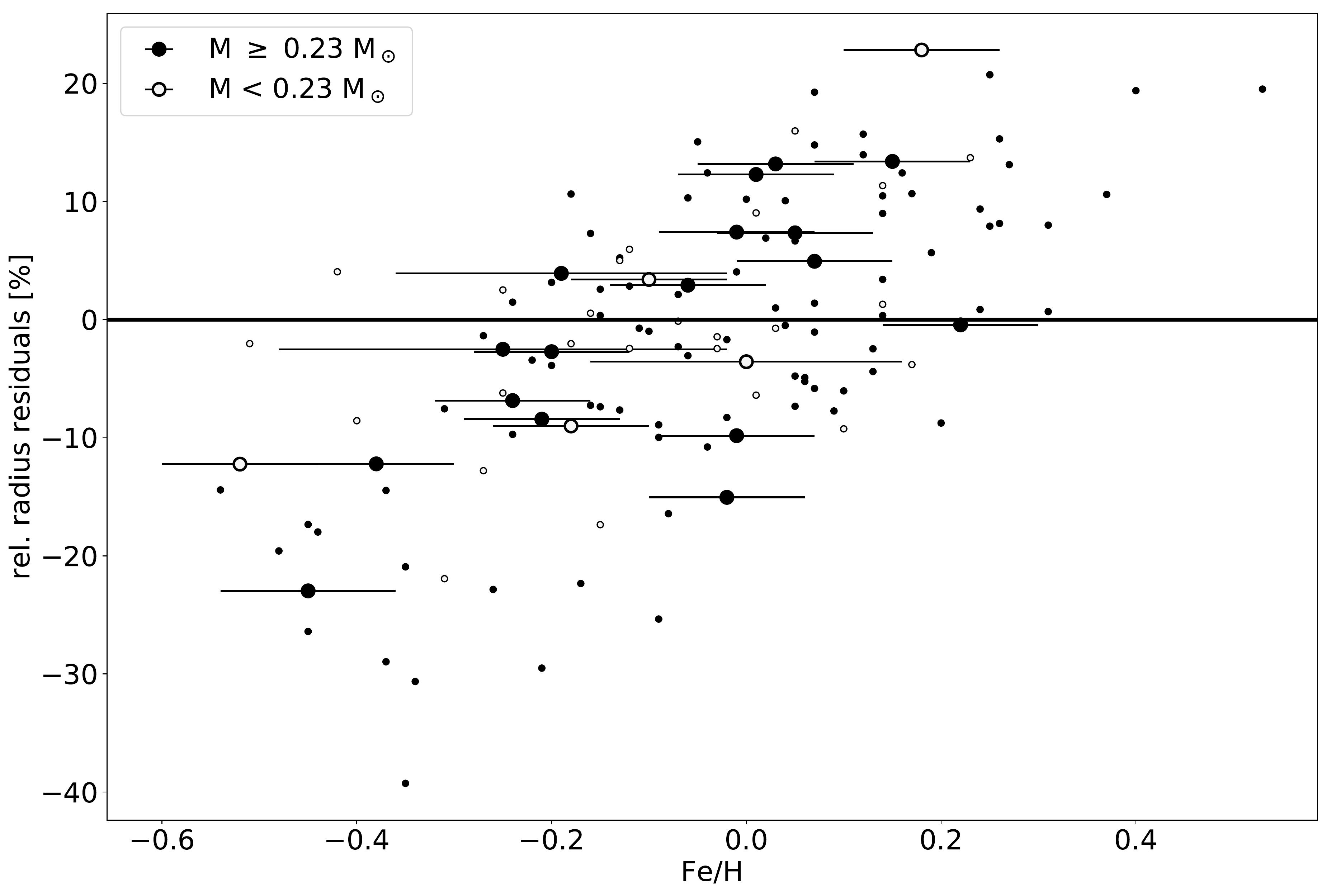}
\caption{ Residuals after subtracting Eq. \ref{eq:RTeff} from the radius measurement as function of stellar metallicity. The Pearson correlation coefficient is $r=0.69$ for $\mstar \geq 0.23\,\msun$ and $r=0.51$ for $\mstar < 0.23\,\msun$. Horizontal line represents the case of zero residuals. (Details are discussed in Sect. \ref{sec:discussion})
\label{fig:res_teffR}}
\end{figure*}

To the \teff-\rstar{} data we fitted two linear polynomials depending on the mass range, namely for stars with $\mstar \geq 0.23\,\msun$ and $\mstar < 0.23\,\msun$. We also tried higher order polynomials, but found in both cases that the higher order coefficients were consistent with zero. We conclude therefore, that for the two cases, the data are best described with two linear polynomials of the form 

\begin{equation}
   \frac\rstar\rsun = 
\begin{cases}
    -1.223(\pm 0.085) + 2.700(\pm 0.138)\,\frac\teff\teffsun\\
    \qquad \text{for  $\mstar \geq 0.23\,\msun$},  \\
    -0.277(\pm 0.060) + 0.869(\pm 0.113)\,\frac\teff\teffsun\\	 
    \qquad \text{for $\mstar < 0.23\,\msun$}.
\end{cases}
\label{eq:RTeff}
\end{equation}
The standard deviation of the residuals are $0.051\,\rsun$ for $\mstar \geq 0.23\,\msun$ and $0.016\,\rsun$ for $\mstar < 0.23\,\msun$ and the respective MADs are $0.033\,\rsun$ and $0.0107\,\rsun$.
In Figure~\ref{fig:res_teffR} we show the residuals after subtracting Equation~\ref{eq:RTeff} as a function of metallicity. The slope in the data indicates a correlation between metallicity and radius, hence, we calculated the Pearson's correlation coefficient ($r$). For stars with $\msun \geq 0.23$ we get $r=0.69$ and for $\msun < 0.23$ $r=0.51$, respectively. We found that stars with higher metallicity have slightly lager radii, and sub-solar metallicity stars lower radii. This correlation is strong for partially convective stars and moderate for fully convective ones. \citet{burrows:2007} proposed that enhanced opacity in atmospheres due to enhanced metallicity could cause inflated radii in giant planets. Given that we find a correlation between metallicity and radius, it is possible to have a similar effect in \Mds. The metallicity effect on the radius can be best described by two linear polynomials of the form:
\begin{equation}
    \frac{\Delta\rstar}{\rstar} = 
\begin{cases}
-0.0060(\pm 0.0093) + 0.4166(\pm 0.0462)\,\feh\\ 
    \qquad  \text{ for  $\mstar \geq 0.23\,\msun$},  \\
0.0187(\pm 0.0176) + 0.2504(\pm 0.0778)\,\feh\\ 
    \qquad  \text{ for $\mstar < 0.23\,\msun$}.
\end{cases}
\label{eq:R_Feh}
\end{equation}

\begin{figure*}
\includegraphics[scale=0.47]{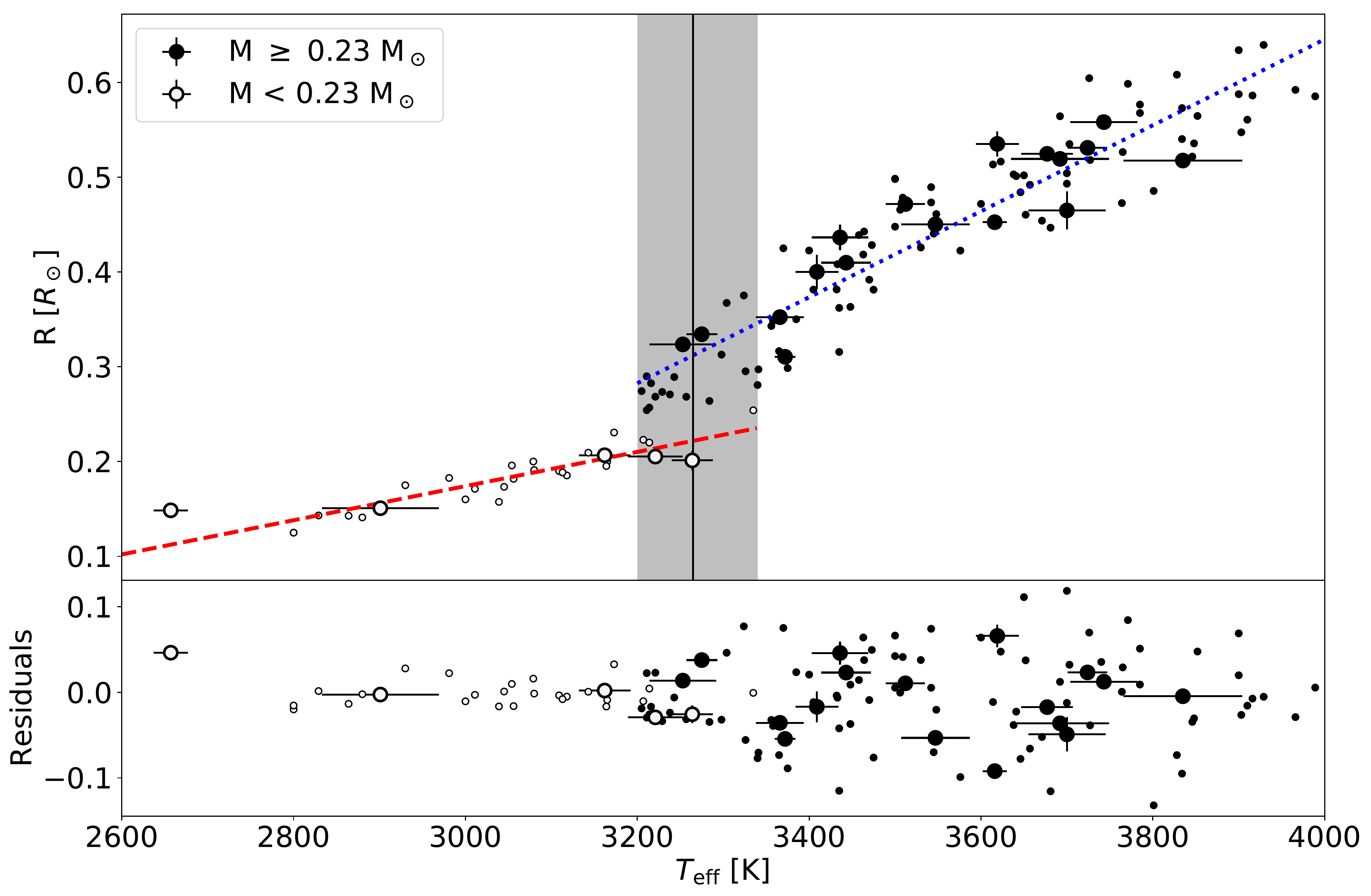}
\caption{ Same as Fig. \ref{fig:teffR}, but correcting the radius for possible metallicity effects. (Details are discussed in Sect. \ref{sec:discussion})
\label{fig:teffR_corrected}}
\end{figure*}

In Fig. \ref{fig:teffR_corrected} we show \teff-\rstar{}, where we corrected the stellar radius for possible metallicity effects using Eq. \ref{eq:R_Feh}. The best-fit polynomials in this case are:

\begin{equation}
    \frac\rstar\rsun = 
\begin{cases}
-1.169(\pm 0.063) + 2.620(\pm 0.103)\,\frac\teff\teffsun\\ 
    \qquad \text{for  $\mstar \geq 0.23\,\msun$},  \\
-0.367(\pm 0.050) + 1.041(\pm 0.094)\,\frac\teff\teffsun\\ 	 
    \qquad \text{for $\mstar < 0.23\,\msun$}.
\end{cases}
\label{eq:RTeff_corrected}
\end{equation}
The standard deviation of the residuals are $0.038\,\rsun$ and $0.013\,\rsun$ respectively, which is slightly lower compared to neglecting the influence of metallicity on the radius. The median absolute deviations of the residuals are $0.029\,\rsun$ and $0.008\,\rsun$.
 
\begin{figure*}
\includegraphics[scale=0.47]{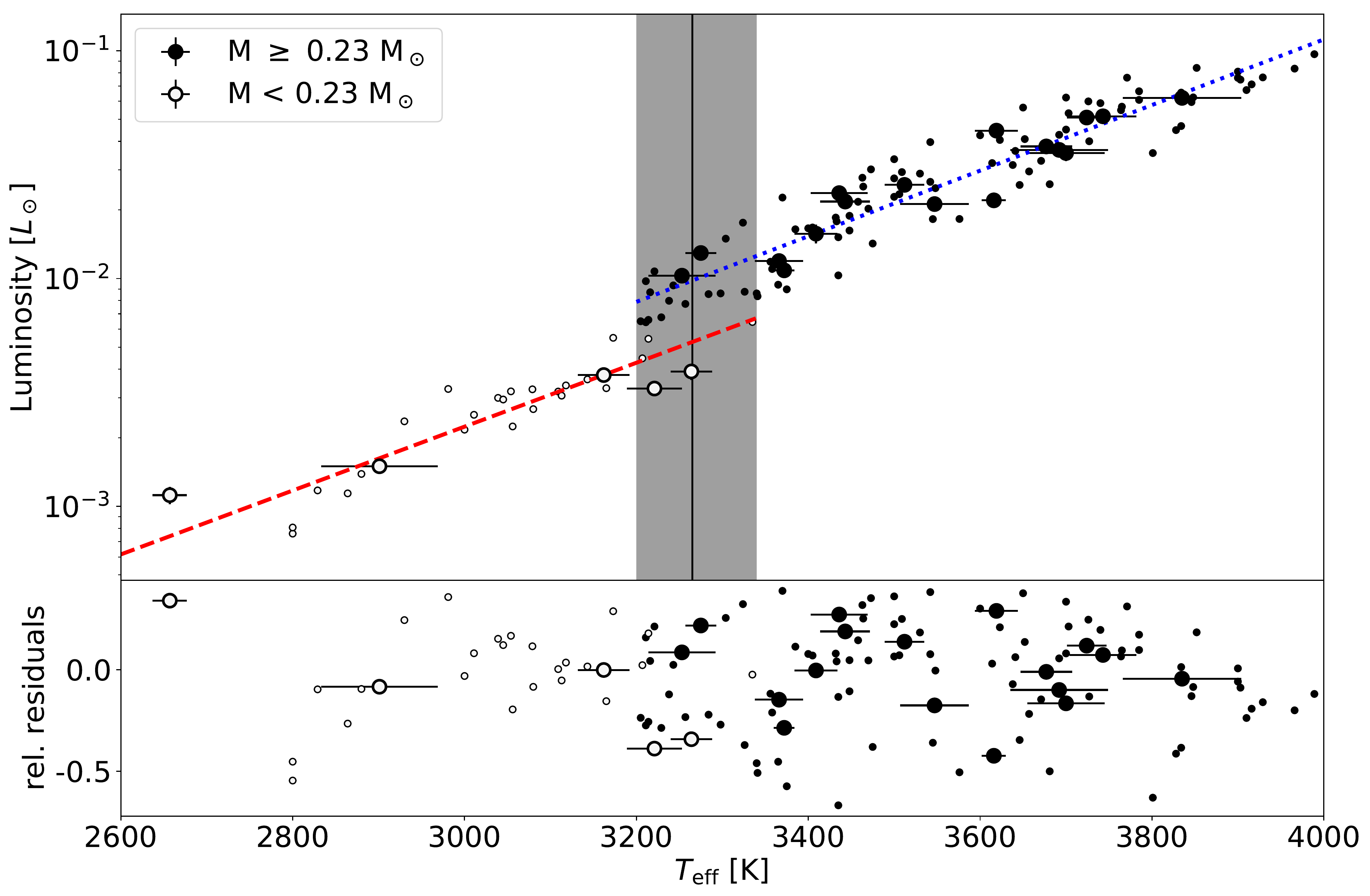}
\caption{Empirical HR diagram  for two different stellar mass populations (filled and empty circles respectively). The red dashed line shows our fitted polynomials for stars with masses $< 0.23\,\msun$ and blue dotted one for masses $\geq 0.23\,\msun$). We identify a discontinuity (grey shaded area) reflecting the transition region between partially and fully convective stars. Our sample is depicted by the large dots. The sample from \citet{mann:2015} is represented by the smaller dots, which also have their error bars suppressed. Lower plot shows the relative residuals after subtracting Eq. \ref{eq:RLum} from the luminosity value (\ensuremath{ \frac{\rm calc.~luminosity - polynomial}{\rm calc.~luminosity} }). (Details are discussed in Sect. \ref{sec:discussion})
\label{fig:lumteff}}
\end{figure*}

Based on our observations and our inferred physical parameters, we further show in Figure~\ref{fig:lumteff} the empirical HR-diagram for the two different mass populations. We also can identify a transition region in the HR-diagram. We establish the following linear ($\log$ \lstar)-\teff-relation for the two different populations

\begin{equation}
   \log \lstar = 
\begin{cases}
-6.710(\pm 0.179) + 8.318(\pm 0.290)\,\frac\teff\teffsun\\
      	\qquad \text{for $\mstar \geq 0.23\,\msun$}, \\
-6.856(\pm 0.345) + 8.099(\pm 0.653)\,\frac\teff\teffsun\\
        \qquad \text{for $\mstar < 0.23\,\msun$}.\\
\end{cases}
\label{eq:RLum}
\end{equation}

\subsection{Transition into the fully convective regime\label{subsec:transition}}

Theoretical stellar evolution predict a transition from partially convective stars into the fully convective stellar regime to occur at stellar mass somewhere between 0.2\,\msun{} \citep{dorman1989} and 0.35\,\msun{} \citep{chabrier:1997:stellarstructure}, depending on the underlying stellar model. While a partially convective star still resembles a sun-like structure, having a radiative zone and a convective envelope, fully convective stars have no such zone. Our observations indicate that the limit between partially and fully convective regime is around $\approx 0.23\,\msun$ and between 3200 and 3340\,K. The lack of a detection of this transition in previous works can be explained mainly by the fact that very few single \Mds{} with temperatures below 3270\,K have interferometrically measured radii. In fact, \citet{boyajian:2012:mdwarfs} shows only two \Mds{} with temperatures below this value. Moreover, they include in their work mostly one of the stars (GJ~699) which is in the fully convective regime, as GJ~551 was excluded from most of their analyses. Another reason is that previous radius measurements of fully convective stars relied on eclipsing \Md{} binaries, where the disentanglement of the respective components is not straightforward. Finally, most radius estimates rely on stellar evolution models rather than direct measurements, i.e.\ in many cases the radius has not been measured directly.

Furthermore, we find that the linear term of the polynomial in Equation~\ref{eq:RTeff} shows a steeper slope for stars with $\mstar \geq 0.23\,\msun$, than for $\mstar < 0.23\,\msun$. This is possibly due the fact that stars with $\mstar \geq 0.23\,\msun$ still have a radiative zone which decreases with shrinking \teff. For \Mds{} with $\mstar < 0.23\,\msun$ the stars are fully convective, i.e.\ the convective zone extents towards the core. Therefore, the linear term for \Mds{} with masses below 0.23\,\msun{} indicates a more flattened slope. The gentle slope for masses below 0.23\,\msun{} is consistent with the fact that fully convective stars have similar spectral types due to $\mathrm{H}_2$ formation, which also flattens the radius-temperature relation \citep{ChabrierBaraffe:2000}.


\subsection{M-dwarfs in the context of Gaia}
\label{sec:Gaia}
\begin{figure}
\includegraphics[scale=0.23]{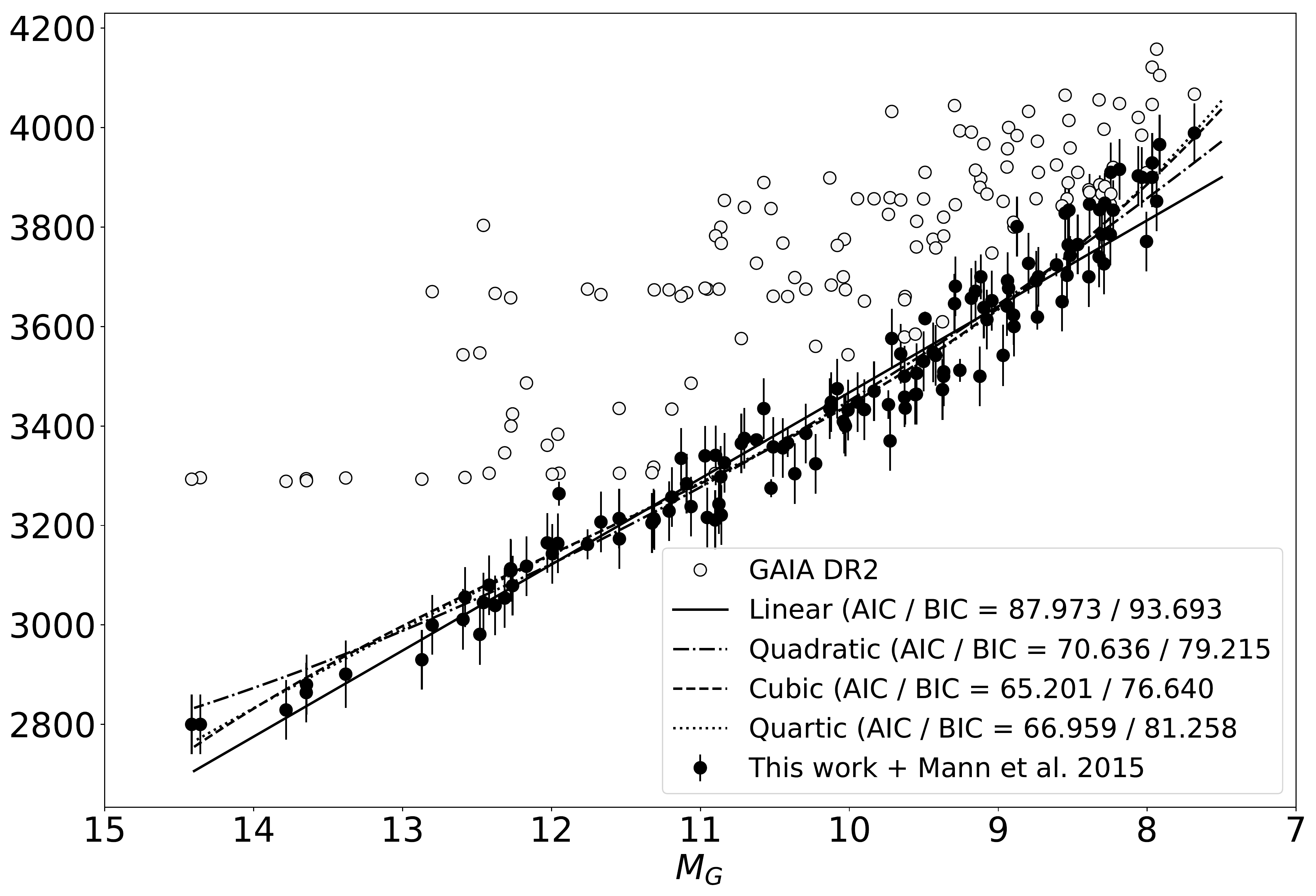}
\caption{Effective temperature as function of absolute Gaia G magnitude. Empty circles show \teff\ estimated by Gaia DR2 Apsis. Filled dots show our measurements and the ones from \citet[]{mann:2015}. Lines show polynomial fits of different orders. Their respective AIC and BIC values are given in the legend. (Details are discussed in Sect. \ref{sec:Gaia})
\label{fig:Gmag_Teff}}
\end{figure}

In Sect. \ref{subsec:multinest} we noticed a considerable difference between \teff\ for M-dwarfs inferred from Gaia three band photometry \citep{gaiadr2_teff:2018} and estimates found here and in the literature \citep{neves:2014:feh,mann:2015} . Therefore, we establish an empirical calibration relation for stars with very well measured G magnitudes and parallaxes from Gaia. We use these two measurements to calculate the absolute G magnitude $M_G$, which we relate  to the \teff. In Fig.~\ref{fig:Gmag_Teff} we show \teff\ as a function of $M_G$. The empty circles show the stellar \teff\ as estimated by Gaia Apsis, whereas the filled circles show our \teff\ measurements and the ones from \citet{mann:2015}. The previously shown discrepancy is also visible in Fig.~\ref{fig:Gmag_Teff}. We determine an empirical relation to obtain \teff\ from $M_G$. In our attempt to find the best relation, we fitted polynomials of different degrees and we calculated their respective AIC and BIC, see Fig. \ref{fig:Gmag_Teff}. The best relation is described by a cubic polynomial of the form:

\begin{multline}
\teff = 10171.7(\pm 1449.6) -1493.4(\pm 410.8) M_G \\
+ 114.1(\pm 38.3) M_G^2 -3.2(\pm 1.2) M_G^3 \\
\end{multline}

The standard deviation of the residuals is $53\,\teff$ and the median absolute deviation (MAD) $36\,\teff$. The errors of the polynomial coefficients (closed brackets) are estimated from the covariance matrix. 

\begin{figure}
\includegraphics[scale=0.23]{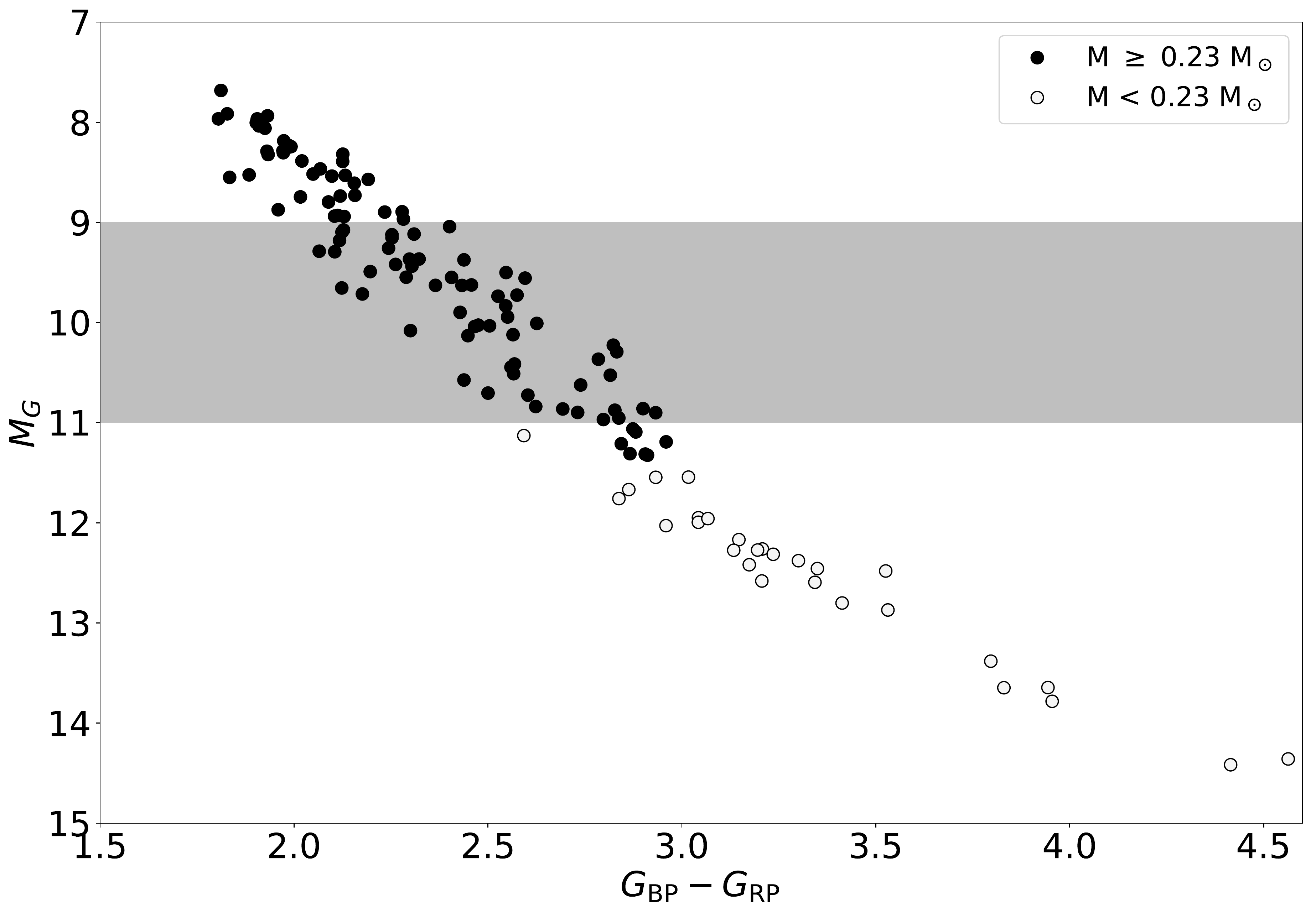}
\caption{Absolute Gaia G magnitude versus $G_{\rm BP} - G_{\rm RP}$ for M-dwarfs with different masses. Gray shaded area shows the region, where \citet{jao:2018} found a gap in their HR-diagram. (Details are discussed in Sect. \ref{sec:Gaia})
\label{fig:G_BP_RP}}
\end{figure}

Recently, \citet{jao:2018} presented an investigation showing a $\sim$0.05 mag gap in the HR diagram constructed from M-dwarfs using the Gaia DR2. The authors attributed this gap to a possible transition from partially to fully convective low-mass stars. However, recent simulations by \citet{MacDonald:2018} argued that this gap can be explained by $^{3}$He instabilities of low-mass stars rather than the before mentioned transition region. This $^{3}$He instabilities are caused by stars with a thin radiative zone, slightly above the transition to fully convective stars. These instabilities can produce energy fluctuations and a dip in the luminosity function \citep{vansaders:He:2012,MacDonald:2018}. In Fig. \ref{fig:G_BP_RP} we show $M_G$ over $G_{BP}-G_{RP}$ and mark the region where \citet{jao:2018} found their discontinuity (grey shaded area). The locus of our discovered discontinuity is slightly below the one from \citet{jao:2018}. This increases the likelihood of the finding from \citet{MacDonald:2018} and our claim having observed the transition region between fully and partially convective stars, which should occur slightly below the $^{3}$He instability region. 

\section{Conclusion} \label{sec:conclusion}

We have measured physical parameters of 13 M-stars covering the partially and fully convective regime using interferometric measurements from the VLTI and parallaxes from Gaia DR2. Our measurements extend to lower \teff{} than previous interferometric studies, and we use them augmented with literature data to present improved empirical relations between stellar radius and mass, and between stellar radius and luminosity as a function of \teff{}.
Analysing residuals to our relations, we identified a general trend that late-type stars with higher metallicity are slightly inflated, whereas for stars with lower metallicity we measure predominantly smaller radii. We find this correlation to be strong for stars with $\mstar \ge 0.23\,\msun$ and moderate for $\mstar < 0.23\,\msun$, respectively. We also found that Gaia \teff{} values are significantly underestimated ($\approx 8\%$) for M-dwarfs.

The most striking feature we identified in our data is a sharp transition in the relation between \rstar{} and \teff{}, as well as in the empirical HR diagram, which we identify as reflecting the transition between partially and fully convective stars. While previously only a hint for this change had been inferred indirectly, we now have a possible direct observation. We showed that this change happens at $\sim 0.23\,\msun$ and between 3200 and 3340 K. In this region we measure radii in the range from 0.18 to 0.42\,\rsun. Thus, our findings put strong constraints on the stellar evolution and interior structure models. 

\section*{Acknowledgements}
We thank the reviewer for their helpful comments
on the manuscript.
M.R. acknowledges support from CONICYT project Basal AFB-170002.
Partially based on observations obtained via ESO under program IDs 090.D-0917, 091.D-0584, 092.D-0647, 093.D-0471.
A.J.\ acknowledges support from FONDECYT project 1171208, CONICYT project BASAL AFB-170002, and by the Ministry for the Economy, Development, and Tourism's Programa Iniciativa Cient\'{i}fica Milenio through grant IC\,120009, awarded to the Millennium Institute of Astrophysics (MAS). R.B.\ acknowledges additional support from project IC120009 ``Millenium Institute of Astrophysics  (MAS)'' of the Millennium Science Initiative, Chilean Ministry of Economy.  
This work made use of the Smithsonian/NASA Astrophysics Data System (ADS) and of the Centre de Donn\'ees astronomiques de Strasbourg (CDS). This research made use of Astropy, a community-developed core Python package for Astronomy (Astropy Collaboration, 2013).
This work has made use of data from the European Space Agency (ESA) mission
{\it Gaia} (\url{https://www.cosmos.esa.int/gaia}), processed by the {\it Gaia}
Data Processing and Analysis Consortium (DPAC,
\url{https://www.cosmos.esa.int/web/gaia/dpac/consortium}). Funding for the DPAC
has been provided by national institutions, in particular the institutions
participating in the {\it Gaia} Multilateral Agreement.




\bibliographystyle{mnras}




\appendix



\bsp	
\label{lastpage}
\end{document}